**Pulsed sextupole injection for Beijing Advanced Photon Source with ultralow emittance**

JIAO Yi (焦毅), XU Gang (徐刚)

Institute of high energy physics, CAS, Beijing 100049, China

**Abstract**: In this paper we present the physical design of the pulsed sextupole injection system for Beijing Advanced Photon Source (BAPS) with an ultralow emittance. The BAPS ring lattice is designed in such a way that two options of pulsed sextupole injection are allowed, i.e., with septum and pulsed sextupole in different drift spaces or in the same drift space. We give the magnetic parameters of the injection system and the optimal condition of the optical functions for both options. In addition, we find that the pulsed sextupole induces position-dependent dispersive effect and causes non-ignorable effect on the injection efficiency in a storage ring with a relatively small acceptance, which should be well considered.

1. **Introduction**

"Ultimate" storage rings (USRs) [1], with equal transverse emittances at diffraction limit for X-rays of interest for user community and with much higher performance than existing rings, have been extensively studied in the past few years [2-7]. It is believed that USRs can provide highly stable photon beams having low peak brightness with high average brightness and high pulse repetition rate, photons that do not over-excite or damage samples the way those from free electron lasers do, and they can serve a large number of diverse users simultaneously. However, associated with the reduced emittance and demanding lattices, several technical challenges exist for the actual implementation of USRs, one of which is to inject the electron beam into the storage ring with high efficiency. The proposed methods include on-axis "swap-out" injection [8] and off-axis injection with pulsed multipole magnets [9-10]. In this paper, taking the Beijing Advanced Photon Source (BAPS) as example, we thoroughly discuss the latter scheme, i.e., injection with a pulsed sextupole.

Scientists at the Photon Factory (PF) proposed and demonstrated the pulsed sextupole injection in 2010 [10]. Experiment shows that the residual oscillation of the stored beam during "top-up" injection with a pulsed sextupole is much smaller than that with conventional bump kickers. Because of the superiority of the pulsed sextupole injection, many institutions worldwide have considered such injection scheme to improve the machine performance [11-13].

The schematic view of the pulsed sextupole injection is shown in Fig. 1. The injected bunch from the transfer line is deflected into the ring by a septum magnet. After passing a part of the beam line, the bunch enters the pulsed sextupole with coordinate of $(x, x')$, where $0 < |x| < min(x_{ph}, x_{dy})$, with $x_{ph}$ and $x_{dy}$ being the horizontal physical and dynamic aperture, respectively. Provided the normalized integral sextupole field strength $K_2 = B''l/(B\rho)$ satisfies

$$x' + \frac{1}{2}K_2 x^2 = 0, \qquad (1)$$

the injected beam will be captured into the ring, where $B''$ and $l$ are the gradient and the length of the sextupole, respectively; $(B\rho)$ is the magnetic rigidity of the reference particle. By contrast, the stored beam passing the pulsed sextupole through the magnetic center sees approximately zero kick, and thus experiences little disturbance.

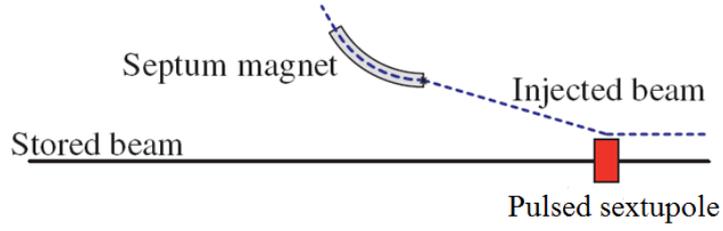

Fig. 1. Schematic view of the pulsed sextupole injection

BAPS, a 5-GeV storage ring-based light source, is planned to be built in Beijing to satisfy the increasing requirements of high-brilliance, high-coherence radiation from the user community [14]. A storage ring with circumference of 1364.8 m and natural emittance of 51 pm is recently designed. We preserve the possibility of further decreasing the emittance to diffraction limited emittances (~10 pm) in both transverse planes by using damping wigglers and locally-round beam production method [15]. Figure 2 shows the lattice layout and the optical functions along the ring. One can see that except 34 standard low-beta sections for high-brightness radiation, there are two high-beta straight sections specifically designed for injection. With delicate optics matching, the phase advance of each injection section ($\psi_{inj,x}$ $\psi_{inj,y}$) is tuned to ($\psi_{sta,x}+2\pi$, $\psi_{sta,y}$), with $\psi_{sta}$ being the phase advance of each standard section, so as to reserve the phase advance between the chromaticity-correction sextupoles in different straight sections and cancel the nonlinear driving terms as much as possible. In addition, high beta function in the middle of the injection section helps to obtain relatively large horizontal dynamic aperture $x_{dy}$. Figure 3 presents the on-momentum horizontal acceptance of the ring in the middle of the injection section obtained with the AT program [16]. One can see that $x_{dy}$ is slightly larger than 10 mm, which hardly satisfies the requirement of the conventional bump injection, but promises pulsed sextupole injection. On the other hand, high beta function results in relatively small acceptance of the horizontal angular deviation (on the scale of 0.1 mrad) which, as will be shown below, leads to difficulty in the optimization of the injection efficiency.

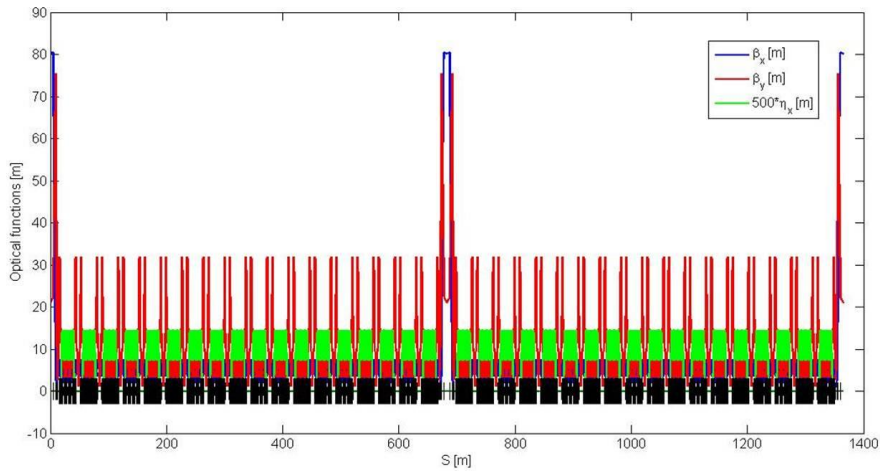

Fig. 2. Lattice layout and the optical functions for one design of the BAPS storage ring.

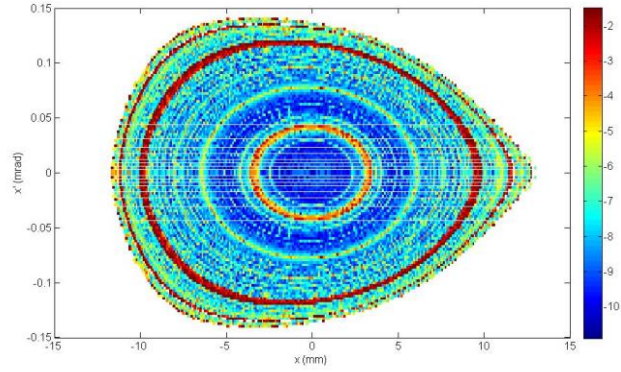

Fig. 3. (color online) On-momentum horizontal acceptance in the middle of the injection section of the BAPS storage ring, obtained with numerical tracking of 1024 turns. The colors, from blue to red, represent the stabilities of the particle motion, from stable to unstable.

For BAPS, it is assumed that the beam is injected from a 400-m booster, with natural emittance $\varepsilon_{x0}$ of 4 nm and rms energy spread $\sigma_\delta$ of 0.1%. In this paper, the end of the transfer line, i.e., the end of the septum magnet is referred to as the injection point (IP), and the end of the pulsed sextupole magnet is denoted as PSM. Since there are ample drift spaces in the injection section, two options of pulsed sextupole injection are allowed. As shown in Fig.4, the septum magnet and the pulsed sextupole can be placed in different drift spaces, or in the same drift space. We optimize the injection efficiency for both options, during which we find that the position-dependent dispersion functions induced by the pulsed sextupole induces (mentioned in Ref. [13]), have non-ignorable influence on the injection for BAPS.

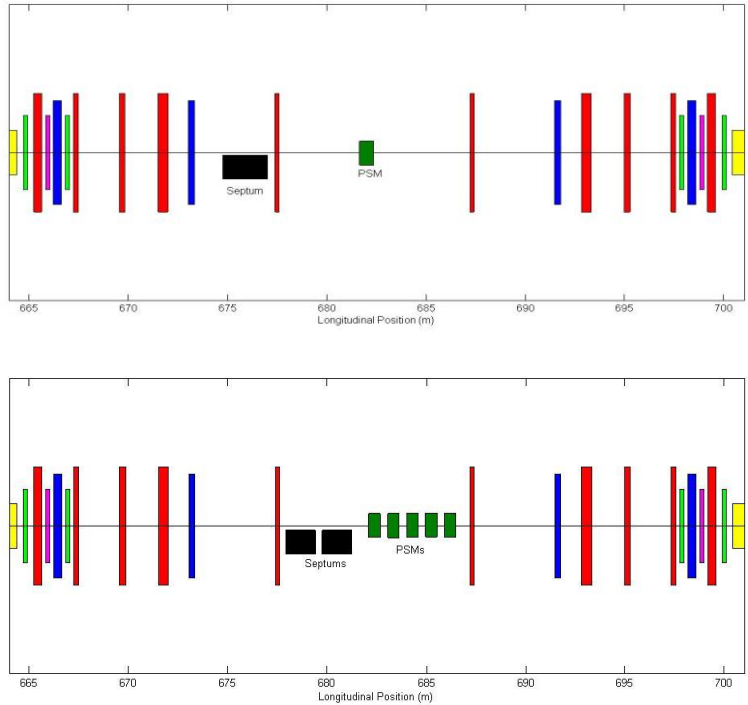

Fig. 4 Schematic view of two options of the pulsed sextupole injection in BAPS storage ring, the first option (upper plot) with septum and pulsed sextupole located in different drift spaces, while the second option (lower plot) with the magnets in the same drift space. The abbreviation "PSM" in the figure indicates the pulsed sextupole magnet.

Unlike the second option, there are focusing quadrupole located between the septum and the pulsed sextupole in the first option. Therefore the phase advance between the IP and the PSM is larger, resulting in smaller sextupole strength and higher injection efficiency. For the same reason the first injection option is, however, mode-dependent, i.e., the optimal solution for the injection probably varies with the change of the ring optics. Since the optimization processes for the two options are similar, we will present the detailed analysis for the first option, while show only the main results for the second.

The paper is arranged as follows. In Sec. 2, the injection system is designed by considering the injected beam as a "point" charge (with zero emittance), where we determine the magnetic parameters of the injection system and the optimal dispersion functions at the IP. In Sec. 3, the injection efficiency is optimized by considering the injected beam with finite emittance, where we study the position-dependent dispersive effect induced by the pulsed sextupole and derive the optimal Courant-Snyder parameters at the IP. Finally conclusions are given in Sec. 4.

## 2. Injection system design by considering beam with zero emittance

In the first option of the pulsed sextupole injection, a 2-m septum magnet is located in a 4-m drift space, with a distance of 0.5 m from the downstream closest quadrupole; and the end of the pulsed sextupole is at the center of the next 9.6-m drift space (see Fig. 4). The position of the injected bunch relative to the stored beam at the IP is shown in Fig. 5. The blade is at -9 mm from the trajectory of the unperturbed stored beam, with a thickness of 2.5 mm. The injected bunch at the IP is at $x_{inj,c}$ = -14 mm, where the subscript "$c$" indicates the center of the beam and will be ignored in the following text for simplification. In this section, we will always consider only the central orbit of the injected bunch. In other words, the injected bunch is treated as a "point" charge and the finite-emittance related effect is not taken into account.

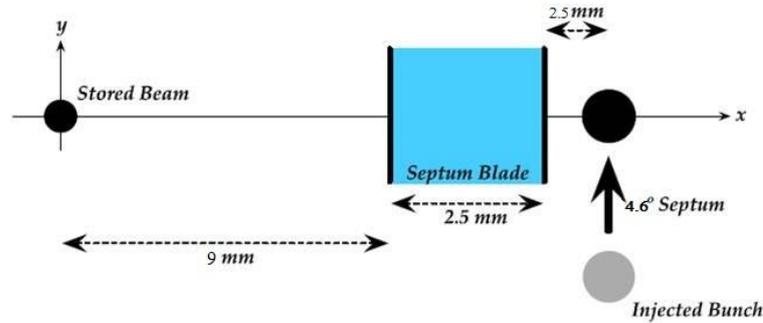

Fig. 5. Schematic view of the position of the injected bunch relative to the stored beam at the IP.

The injection system is modeled as below. The injected bunch starts from the IP with coordinate of $X_{IP,0} = (x_{IP,0}, x'_{IP,0}, \delta_{IP,0})^T = $ (-14 mm, $x'_{IP,0}$, 0)$^T$, where the superscript "T" means the transpose of a vector (or a matrix), the subscript "0" indicates zero momentum deviation, and the variables unrelated to the injection are ignored. A thin-lens model of the pulsed sextupole is first used to simplify the analysis. For each value of the $x'_{IP,0}$, the injected bunch has a specific trajectory from the IP to the PSM. We record the coordinate at the PSM $X_{psm,0} = (x_{psm,0}, x'_{psm,0}, 0)^T$ if $|x_{psm,0}| \in$ (2, 6) mm, obtain the required PSM strength $K_2$ according to Eq. (1), and calculate the reduced emittance $\varepsilon_{red}$, i.e., the area surrounded by the motion started from ($x_{psm}$, 0) in the horizontal phase space,

$$\varepsilon_{red} = (1+\alpha_{psm}^2)x_{psm}^2 / \beta_{psm}, \tag{2}$$

where $\alpha_{psm}$ and $\beta_{psm}$ are the Courant-Snyder parameters at the PSM.

The calculation results are summarized in Fig. 6. It shows that to reduce the required sextupole strength $K_2$, a negative $x_{psm}$ with large absolute value is preferred. On the other hand, considering ring acceptance reduction due to magnetic errors and other factors in the actual implementation, choosing an $x_{psm}$ close to the edge of the ring acceptance is not recommended. As a compromise, the solution corresponding to a moderate $x_{psm} \sim$ -5 mm is chosen, i.e., $x'_{IP,0}$ = -0.79 mard, $K_2$ = 154.7 and $\varepsilon_{red}$ = 0.31 μm.

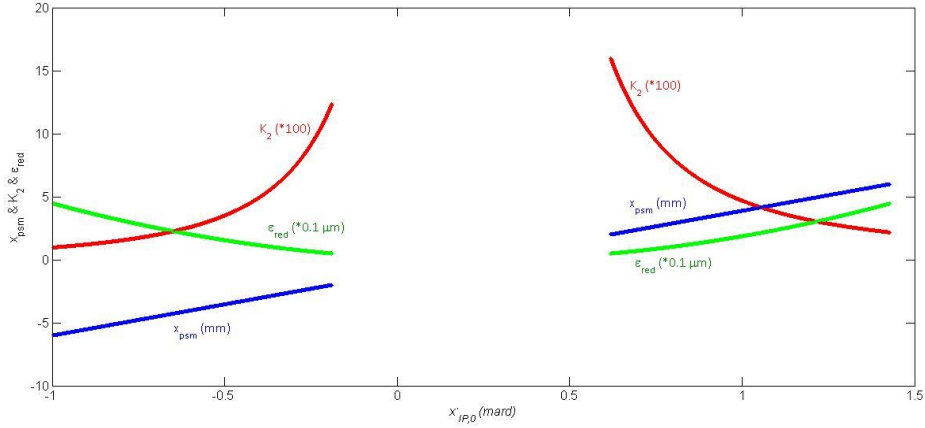

Fig. 6. Variation of the horizontal offset at the PSM $x_{psm}$, the required sextupole strength $K_2$, and the reduced emittance $\varepsilon_{red}$ with the central horizontal angular deviation of the injected bunch $x'_{IP,0}$ for the first injection option.

Associated with the nonlinear field, the trajectory of a particle in a finite-length sextupole is different from that in a thin-lens sextupole. In our design a MAX-IV-type pulsed sextupole [13] is adopted in the first option, with bore diameter of 36 mm. According to the available maximum pole face field (~ 0.6 T) and the obtained $K_2$, the sextupole length is chosen to be 0.7 m. We repeat the above analysis while replacing the thin-lens model with this 0.7-m pulsed sextupole, and finally obtain the exact solution corresponding to $x_{psm}$ = -5 mm, i.e., $x'_{IP,0}$ = -0.64 mard and $K_2$ =151.2.

With the found parameters, the transportation of the injected on-momentum particle can be described in the form

$$X_{psm,0} = \mathcal{M}_{I2P} X_{IP,0}, \tag{3}$$

where $\mathcal{M}_{I2P}$ refers to the transfer map from the IP to the PSM.

Furthermore, we need to make sure that the pulsed sextupole can capture the off-momentum injected particles which start from the IP with coordinate of $X_{IP,\delta}$ = (-14 mm, -0.64 mrad, δ)$^T$,

$$X_{psm,\delta} = \mathcal{M}_{I2P} X_{IP, \delta}, \tag{4}$$

Simulation shows that there exists dispersive effect from the IP to the end of the PSM,

$$\begin{aligned} x_{psm,\delta} &= x_{psm,0} + A\delta, \\ x'_{psm,\delta} &= x'_{psm,0} + B\delta, \end{aligned} \tag{5}$$

where $A = -0.012$, $B = -0.008$. For particle with large momentum deviation $\delta \sim 3\sigma_\delta = 0.3\%$, it causes an $x'$-deviation of 0.024 mard compared with the on-momentum injected particle.

This dispersive effect can be corrected, if the dispersion functions $(D_x, D_x')$ at the IP is set to such values that the dispersion functions seen by the injected particles at the PSM are zero. To this end, we set the particle's coordinate at the PSM to $X_{psm,\delta} = (x_{psm,0}, x'_{psm,0}, \delta)^T$ and trace back from the PSM to the IP,

$$X_{IP,\delta} = \mathcal{M}_{P2I} X_{psm,\delta}, \tag{6}$$

where $\mathcal{M}_{P2I}$ refers to the transfer map from the PSM to the IP.

By comparing the coordinates of the on- and off-momentum particles, the required dispersion functions at IP can be calculated, $(D_x, D_x') = (0.0088 \text{ m}, 0.0017)$. Once these dispersion functions are achieved by adjusting the optics of the transfer line, the off-momentum particle will have the same coordinate as the on-momentum particle at the PSM.

Up to now the magnetic parameters as well as the dispersion condition at the IP have been fixed. These results for both options are summarized in Table 1.

Table 1: Parameters and dispersion condition for the pulsed sextupole injection in BAPS

| Parameter | First option | Second option | Unit |
|---|---|---|---|
| No. of septum magnets | 1 | 2 | |
| Septum magnet length | 2 | 1, 1 | m |
| Septum bending angle | 80 | 30, 70 | mard |
| No. of pulsed sextupoles | 1 | 5 | |
| Pulsed sextupole length | 0.7 | 0.6, 0.6, 0.6, 0.6, 0.6 | m |
| Pulsed sextupole bore diameter | 36 | 48 | mm |
| Pulsed sextupole gradient $B''$ | 3602 | 1000 | T/m$^2$ |
| Central horizontal offset of the injected beam at the PSM $x_{psm}$ | -5 | -5 | mm |
| Dispersion function at the IP $D_x$ | 0.0088 | 0.0332 | m |
| Derivative of dispersion at the IP $D_x'$ | 0.0017 | 0.0074 | |

### 3. Injection system design by considering beam with finite emittance

In this section we will consider more actual circumstance, an injected bunch with a finite emittance, and look for the optimal condition for the maximum injection efficiency.

Associated with the limited ring acceptance, especially in the direction of the horizontal angular deviation, special attention is paid on the beam distribution at the PSM. Similar to what we have done in Sec. 2, we first generate a beam distribution with horizontal emittance of 4 nm and with zero momentum deviation at the PSM [denoted as distribution **A**, with the center of $(x_{psm,0}, x'_{psm,0}) = (-5 \text{ mm}, 0)$] that can be mostly captured in the ring acceptance, and then trace back to the IP location.

For simplification, the generated distribution **A** is up-right ellipse in the horizontal phase space. The corresponding second order moments beam matrix is in the form

$$\Sigma_{psm,0} = \begin{pmatrix} \sigma_{x,psm}^2 & 0 \\ 0 & \sigma_{x',psm}^2 \end{pmatrix}, \tag{7}$$

where $\sigma_{x,psm}$ and $\sigma_{x',psm}$ are the horizontal rms beam size and the horizontal rms angular spread, respectively. Whatever the $\sigma_{x,psm}$ is, the horizontal emittance $\varepsilon_{x,psm} = \sigma_{x,psm}\sigma_{x',psm}$ is kept to 4 nm.

Through numerical tracking with the AT program, we get the beam distribution at the IP (denoted as distribution **B**), and then statistically obtain the corresponding second order moments beam matrix,

$$\Sigma_{IP,0} = \begin{pmatrix} <x_{IP,0}^2> & <x_{IP,0}x'_{IP,0}> \\ <x_{IP,0}x'_{IP,0}> & <x'^2_{IP,0}> \end{pmatrix}, \tag{8}$$

where $<Y>$ indicates the square of the standard deviation of the data Y.

The beam matrix can be written in terms of the Courant-Snyder parameters,

$$\Sigma_{IP,0} = \begin{pmatrix} \varepsilon_{x,IP}\beta_{IP} & -\varepsilon_{x,IP}\alpha_{IP} \\ -\varepsilon_{x,IP}\alpha_{IP} & \varepsilon_{x,IP}\gamma_{IP} \end{pmatrix}, \tag{9}$$

where $\alpha_{IP}$ and $\beta_{IP}$ are the Courant-Snyder parameter at the IP, and $\gamma_{IP} = (1+\alpha^2_{IP})/\beta_{IP}$.

From the one-to-one correspondence between the matrix elements in Eq. (8) and those in Eq. (9), the Courant-Snyder parameters at the IP can be calculated. With the obtained $\alpha_{IP}$ and $\beta_{IP}$, we re-generate a beam distribution at the IP [denoted as distribution **C**, the center is $(x_{IP,0}, x'_{IP,0}) = $ (-14 mm, -0.64 mrad) ] with horizontal emittance of 4 nm, and with rms energy spread of 0.1%. Then we track these particles from the IP to the PSM, and record the final beam distribution (denoted as distribution **D**).

Figure 7 shows the results with the $\sigma_{x,psm}$ of the distribution **A** equal to 0.3 mm. One can see that the distributions **C** and **D** are somewhat different from the distributions **B** and **A**, respectively, although the optimal dispersion condition obtained previously is already taken into account. Study shows that this difference is induced by the pulsed sextupole. For the injected particles, the pulsed sextupole acts like a bending magnet with the equivalent bending angle dependent of the horizontal displacement of the injected particle at the PSM, resulting in the position-dependent dispersion functions. As a result, the required dispersion condition at the IP is $x_{psm}$-dependent. Note that the optimal dispersion condition listed in Table 1 is obtained by considering only the central orbit of the injected beam. With this condition, the particles away from the beam center will still experience nonzero dispersive effects. Figure 8 presents the residual dispersion functions in term of the $x_{psm}$ for both options. Compared with the limited acceptance of the BAPS storage ring, the position-dependent dispersion functions will cause a relatively large dilution in the direction of the horizontal angular deviation, especially in the case of the second option. To reduce this effect, a small rms horizontal beam size of the injected beam at the PSM is preferred.

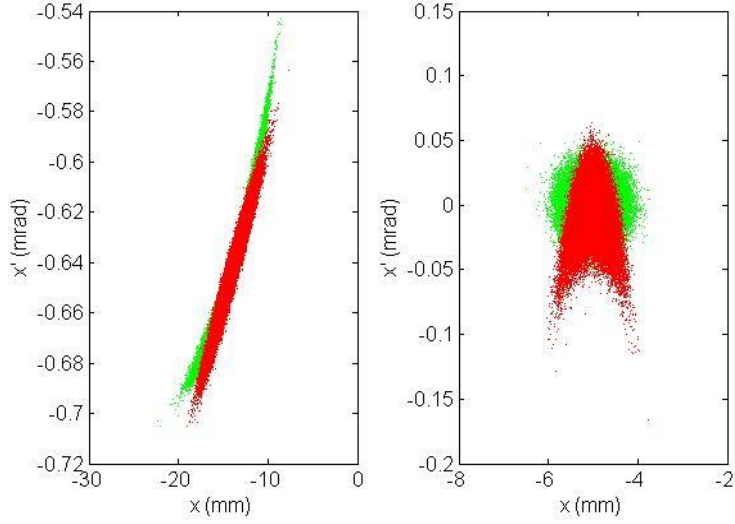

Fig. 7. (color online) Simulation data of the distributions **A** (right plot, green, at the PSM), **B** (left plot, green, at the IP), **C** (left plot, red, at the IP) and **D** (right plot, red, at the PSM), with the $\sigma_{x,psm}$ of the distribution **A** equal to 0.3 mm.

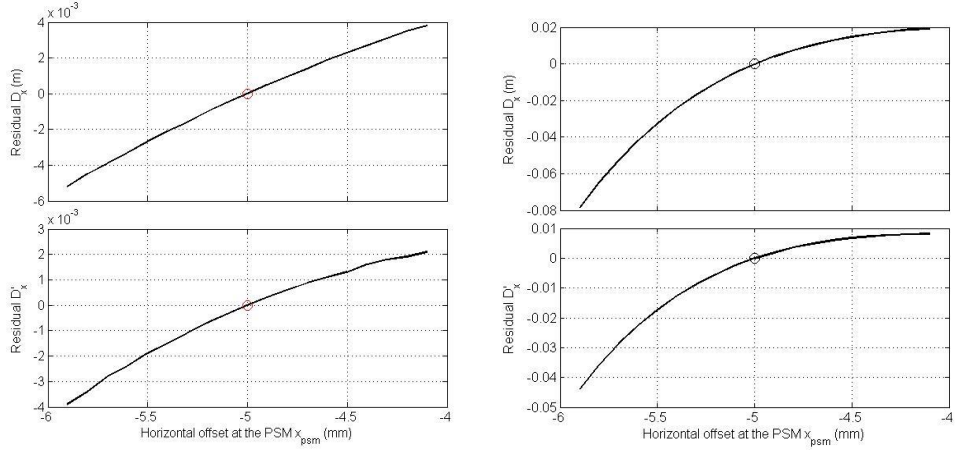

Fig. 8. Residual dispersion functions in term of the $x_{psm}$ for the first (left plot) and the second (right plot) injection option, respectively. The circles indicate the cases with the central horizontal displacement of the injected beam at the PSM.

To optimize the capture rate, as the $\alpha_{IP}$ and $\beta_{IP}$ are obtained for each $\sigma_{x,psm}$ of the distribution **A**, we produce various combinations of $\alpha$ and $\beta$ in the range of ($\alpha_{IP}$-10, $\alpha_{IP}$+10) and ($\beta_{IP}$-10 m, $\beta_{IP}$+10 m), and generate the corresponding distribution **C** for each combination. With numerical tracking we search the optimal combination that result in distribution **D** with the smallest rms angular spread. In addition, since the distribution **D** is not necessarily matched to the acceptance ellipse that is determined by the ring optics, it is put into numerical tracking of at least one damping time (~ 6000 turns) to calculate the relative number of the survived particles, which is defined as the injection efficiency. It is believed that after one damping time the oscillation amplitude of the injected bunch becomes small enough due to radiation damping that all the survived particles will be captured.

Figure 9 shows the variation of the injection efficiency with the $\sigma_{x,psm}$ of the pre-generated distribution **A** for both options. Table 2 summarizes the Courant-Snyder parameters at the IP corresponding to the maximum injection efficiencies for both options.

The maximum injection efficiency of the first option is found to be 99.9%, with the injection and the capture processes (corresponding to point with $\sigma_{x,psm} = 0.15$ mm in the left plot of Fig. 9) illustrated in Fig. 10. The injection and the capture processes of the second option are similar to that of the first option, and are therefore not shown here. Compared to the first option, since stronger pulsed sextupoles are used in the second option, the effect raised from the residual dispersion functions is more significant (see Fig. 8), leading to lower available maximum injection efficiency, ~ 87.3%. However, the second option has the advantage of mode-independency. Thus, it will be very attractive if one can improve the injection efficiency. Study shows that the maximum injection efficiency of the second option increases to about 100% in the case that the horizontal emittance of the injected beam is or below 1 nm, as shown in Fig. 11. Such an emittance can be achieved by re-designing the booster optics to reduce the natural emittance, or by introducing transverse coupling (with solenoids or skew quadrupoles) to the transfer line to reduce the proportion of the horizontal emittance in the natural emittance, or by injecting the beam in the vertical plane instead of the horizontal plane.

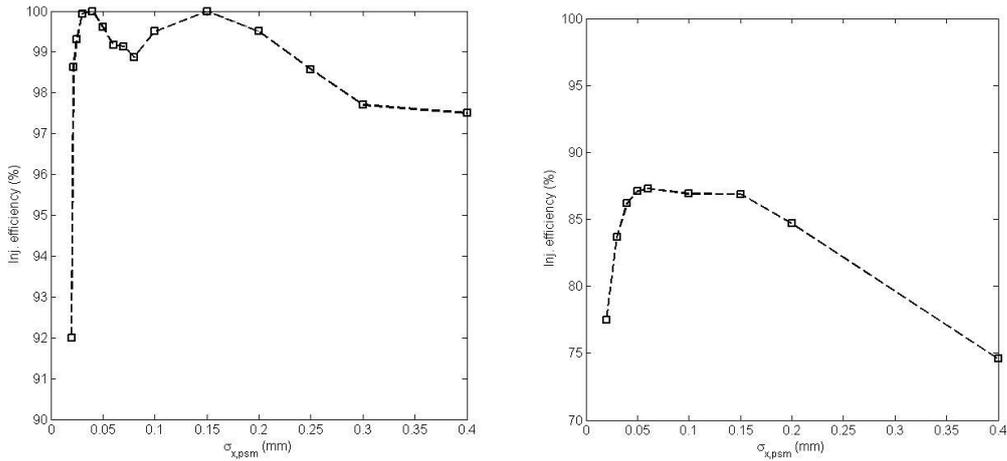

Fig. 9. Variation of the injection efficiency with the $\sigma_{x,psm}$ of the pre-generated distribution **A** for the first (left plot) and the second (right plot) injection option, respectively.

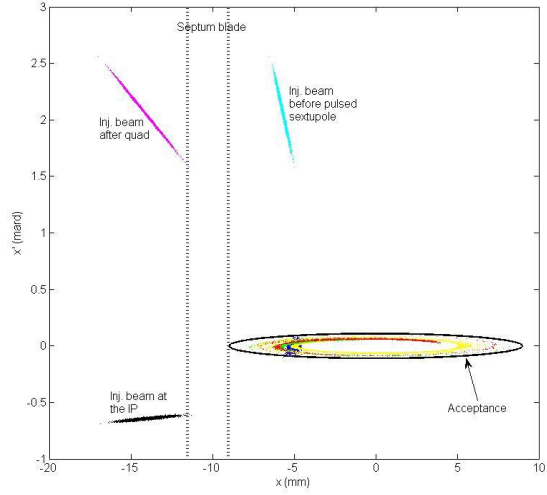

Fig. 10. (color online) Numerical tracking data for the injection and the capture processes with the optimal conditions for the first injection option. The injected beam is modeled using 5000 particles with $\varepsilon_x$ = 4 nm and $\sigma_\delta$ = 0.1%. The first three turns' captured particles are marked with colors of blue, green, red, respectively. The 6000th turn's captured particles are marked with color of yellow.

Table 2: Optimal Courant-Snyder parameters for the pulsed sextupole injection in BAPS

| Parameter | First option | Second option | Unit |
| --- | --- | --- | --- |
| Beta function at the IP $\beta_x$ | 145.7 | 260.8 | m |
| Alfa fucntion at the IP $\alpha_x$ | 2.0 | 96.5 | |

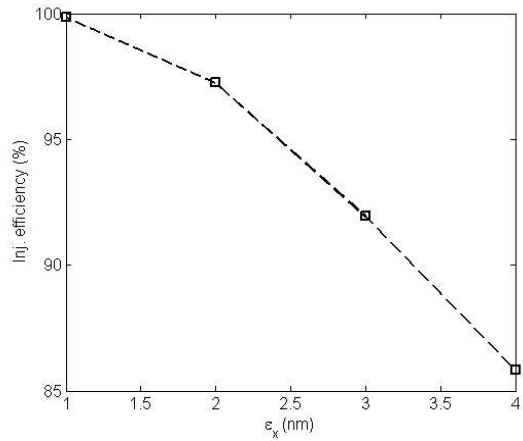

Fig. 11. The available maximum injection efficiency in term of the horizontal emittance of the injected beam for the second injection option.

### 4. Conclusion

In this paper, we present the physical design of the pulsed sextupole injection system for the BAPS storage ring. With optimization of the injection and the capture processes, high injection efficiency is finally obtained. We show that in a storage ring with ultralow emittance and with a relatively small acceptance, the position-dependent dispersive effect induced by the pulsed sextupole may become a problem and should be well considered.

**Acknowledgement**

We appreciate A. Chao, S. Tian, L. Wang and Y. Yuan for helpful discussions. This work was supported by the special fund of the Chinese Academy of Sciences under contract No. H9293110TA.

**References**
[1] Bei M *et al*. Nucl. Instrum. Methods Phys. Res., Sect. A, 2010, **622**: 518
[2] Elleaume P and Ropert A. Nucl. Instrum. Methods Phys. Res., Sect. A, 2003, **500**: 18
[3] Tsumaki K and Kumagai N K, Nucl. Instrum. Methods Phys. Res., Sect. A, 2006, **565**: 394
[4] Borland M. Proc. of the SRI09. Melbourne: AIP, 2009. 911
[5] Jing Y, Lee S Y, Sokol P E. Proc. of the 2011 Particle Accelerator Conference. New York: IEEE, 2011. 781
[6] Cai Y *et al*. Phys. Rev. ST Accel. Beams, 2012, **15**: 054002
[7] XU Gang and JIAO Yi. Towards the ultimate storage ring: the lattice design for Beijing Advanced Photon Source, Chin. Phys. C, to be published
[8] Borland M. Nucl. Instrum. Methods Phys. Res., Sect. A, 2006, **557**: 230
[9] Harada K *et al*. Phys. Rev. ST Accel. Beams, 2007, **10**: 123501
[10] Takaki H *et al*. Phys. Rev. ST Accel. Beams, 2010, **13**: 020705
[11] Sun C *et al*. In Proc.of IPAC10. Kyoto, Japan, 2010, WEPEA068
[12] Resende X R *et al*. In Proc. of IPAC11. Kursaal, Spain, 2011, THPC139
[13] Leemann S C. Phys. Rev. ST Accel. Beams, 2012, **15**: 050705
[14] JIANG Xiaoming *et al*. BAPS preliminary design report, internal report. Beijing: Institute of High Energy Physics, CAS, 2012 (in Chinese)
[15] XU Gang, JIAO Yi and TIAN Saike. Realization of locally-round beam in an ultimate storage ring using solenoids, Chin. Phys. C, to be published
[16] Terebilo A. 2001, SLAC-PUB-8732